\documentclass[12pt]{article}

\def\nextline{\hfill\break}
\def\ds{\displaystyle}
\def\mycomm#1{\nextline\strut\kern-3em{\tt ====> #1}\nextline}
\def\mycommnb#1{\strut\kern-3em{\tt ====> #1}}

\def\eff{{\kern-0.1em{\it eff}\,}}

\usepackage[english]{babel}
\usepackage{epsf}

\def\pslash{\hbox{p\kern-.5em\lower.3ex\hbox{/}}}

\def \be{\begin{equation}} 
\def \ee{\end{equation}} 
\def \bea{\begin{eqnarray}}

\def \eea{ \end{eqnarray}}

\begin{document}

\begin{flushright}
NT@UW-00-1 \\[-.25em]
 SLAC-PUB-8366 \\[-.25em]
TAUP 2620-2000\\[-.25em]
\end{flushright}
\vfill

\begin{center}
\baselineskip=15pt
{{\large \bf Coherent Contributions of Nuclear Mesons to \\[.50em]
 Electroproduction
 and the HERMES Effect}\footnote{This work was supported in part by the United S
tates
Department of Energy under contract numbers DE--AC03--76SF00515,
and DE-FG03-97ER4104 and by a grant from the U.S.-Israel Binational Science Foun
dation.}
 }\\[10 mm]

\vfill

{\bf Gerald A. Miller}\footnote{\tt miller@nucthy.phys.washington.edu }
\\
{\em Physics Department,\\
University of Washington\\
 Seattle, Washington 98195-1560} \\
\vskip 7 mm
{\bf Stanley J. Brodsky}\footnote{{\tt sjbth@slac.stanford.edu}}\\
{\em Stanford Linear Accelerator Center, \\
 Stanford University, Stanford, California 94309}\\[3mm]
and \\
[3mm]
{\bf Marek Karliner}\footnote{\tt marek@proton.tau.ac.il}\\
{\em School of Physics and Astronomy, \\
Raymond and Beverly Sackler Faculty of Exact Sciences\\
Tel Aviv University,
 Tel Aviv, Israel} \\
\end{center}

\vfill
\newpage
\begin{abstract}
\noindent \baselineskip=15pt
 We show that nuclear
$\sigma,\omega$, and
$\pi$ mesons can contribute coherently to enhance the electroproduction
cross section on nuclei for longitudinal virtual photons at low $Q^2$
while depleting the
cross section for transverse photons.  We are able to describe recent HERMES
inelastic lepton-nucleus scattering data at low  $Q^2 $ and small
$x$ using photon-meson and meson-nucleus couplings which are consistent
with (but not determined by) existing constraints  from meson decay widths,
nuclear structure, deep inelastic scattering, and
lepton pair production data.   We find that while
nuclear-coherent pion currents are not important for the present data,
they could be observed at different kinematics. Our model for coherent meson
electroproduction requires the assumption of mesonic currents and couplings
which can be verified in separate experiments.  The observation of
nuclear-coherent mesons in the final state would verify our theory and
allow the identification of a specific dynamical mechanism for higher-twist
processes.
\end{abstract}

\vfill

\renewcommand{\labelenumi}{(\alph{enumi})}
\renewcommand{\labelenumii}{(\roman{enumii})}

\bigskip

Nuclear targets provide a unique way to adiabatically modify
the hadronic environment when testing QCD.  The shadowing and
antishadowing effects of the nuclear medium on the electroproduction cross
section and nuclear structure functions in the Bjorken scaling region are
typically less than 20\%, in qualitative agreement with theoretical
expectations.  However, recent measurements by the HERMES collaboration of the
inelastic lepton-nucleus cross section at low
$Q^2 < 1.5$ GeV$^2$ and small $x < 0.06$ for a $27.5$ GeV positron
beam interacting on gas jet targets at HERA display an extraordinarily
strong nuclear and virtual photon polarization
dependence \cite{Ackerstaff:1999ac}.
The HERMES data for deuterium, $^3$He and
$^{14}$N targets show an anomalously strong nuclear dependence of the
ratio $R^A = \sigma^A_L(x,Q^2)/\sigma^A_T(x,Q^2)$ at low momentum transfer
$Q^2$ and small $x.$ For example,
 $R^{N}/R^{D}$ (nitrogen vs. deuterium) is
$\simeq 5$ at $Q^2 \simeq 0.5$ GeV$^2$, for $x\simeq 0.01.$
This ratio of five  results from a nuclear enhancement of
the longitudinally polarized virtual photoabsorption
by about a factor of 2 and a reduction of the transverse
cross section $\sigma_T(x,Q^2)$ by about a factor of 2.5.
These nuclear effects are very much larger than the typical shadowing effects
mentioned above, and very much larger than previous estimates \cite{raest}
of nuclear enhancements of $R^A$. The nuclear target experiments in this 
kinematic regime
are very challenging because of the large size of the radiative corrections.

It has long been recognized that
a small value of $R$ is the signature of spin 1/2 partons, and, conversely,
a large value would be a signature of bosonic constituents \cite{cagr}.
Furthermore, the rapid decrease of the nuclear
enhancement of $\sigma_L$ with increasing
$Q^2$ seen by HERMES is compatible with elastic scattering of the positron
on a composite bosonic system: the power-law decrease of the square
of the mesonic form factors $\vert F_M(Q^2)\vert^2 \propto 1/Q^4$ could
account for the rapid fall-off of the $R^A/R^D$ enhancement with momentum
transfer.  It is thus natural to interpret the unusual nuclear enhancement of
$R^A$ reported by HERMES in terms of leptons scattering on the mesonic fields
of nuclei.

The fundamental microscopic description of the deep inelastic
electroproduction cross section on nuclei in QCD is based
on quark and gluon degrees of freedom,
but at small $Q^2$ there exist intricate and non-local
correlations which are more readily described in terms of mesonic currents.
A familiar example is the emergence of pions as approximate Goldstone
bosons, which is difficult to see directly in the quark language, but
is immediate in the language of effective Lagrangians.
Thus meson fields are  natural degrees of freedom for the physics of
leptoproduction at low values of $Q^2$ and $x$.

The nuclear enhancement of $\sigma^A_L(x,Q^2)$ for $x < 0.06$ observed by
HERMES suggests constructive interference of amplitudes from mesons emitted by
different nucleons.
The minimum laboratory momentum transfer to the
nucleus in diffractive production of a meson of mass $m_M$ is
$\Delta p_L \simeq (Q^2 + m_M^2) /2 \nu \simeq 2 x_{bj} M_N \approx 60$ MeV,
which is comparable with the inverse nuclear size of nitrogen.
Thus electroproduction can occur from higher-twist subprocesses in which the
lepton scatters elastically on mesons emitted coherently
throughout the entire nuclear volume.

Now let us consider the specific processes which could contribute.  The most
natural effect to consider is positron quasi-elastic scattering from
a nuclear pion.  The emission of a pion leads to a set of low-lying nuclear 
states,
and one could find
significant effects in the sum over states.  However, the probability of finding
a pion at small values of $x$ in the nuclear medium vanishes relative to vector
mesons, reflecting the connection between the Regge behavior of deep
inelastic structure functions and the spin of
the exchanged constituents \cite{Landshoff:1971ff}.
Thus, even though the pion couples strongly to nucleons, we do not expect the
pion to contribute significantly to the HERMES effect at small $x$.  However, as
we discuss below, pionic currents can yield significant effects for
$x\approx 0.25$ and $Q^2 \sim 1$ GeV$^2$.

 Other nuclear mesons besides pions are known to play an important in
nuclear physics.  In simple models \cite{bsjdw},
the exchange of scalar mesons between nucleons leads to
attractive forces which bind the nucleus.  On the other hand, the exchange
of vector mesons supplies the repulsive potential which prevents the collapse
of the nucleus.  The strengths of such
fields in the nucleus are quite large \cite{Furnstahl:1999ff},
corresponding to a significant number of effective bosonic partons at
very small values of $x$ \cite{gam}.
Furthermore, in high energy processes, the effects of
vector mesons are enhanced because of the presence of factors of momentum in
the interaction.  This is an essential feature of the explanation provided here.

Nuclear deep inelastic lepton-scattering and Drell-Yan
experiments place important constraints which limit the
nuclear enhancements of mesons relative to the nucleon;
see the summary \cite{bfs}.  For example, for infinite nuclear matter the
nucleons must carry 90\% of the light cone plus momentum \cite{sick},
which implies mesons can carry no more than 10\%.  Our calculations shall
use models and parameters which respect these constraints.

We begin a quantitative analysis by examining the
consequences of nuclear-coherent vector and scalar mesons.
We find that the process of Fig.~1a in which the interaction between the
virtual photon $\gamma^*$ of momentum $q$ with a nuclear $\omega$ meson which
produces a $\sigma$ meson in the final state can give a significant
contribution to $\sigma_L(A)$.  To understand the shadowing of $\sigma_T$,
we need a process which interferes destructively with the dominant
process (at low $x$) of Fig.~1b.  We find that the process of Fig.~1c in which
the virtual photon converts a nuclear $\sigma$ meson into a vector meson
can supply the necessary destructive interference.  If this amplitude is 1/3 of
the dominant process it leads to a reduction of the cross section of a
factor of 4/9, which is needed to account for the shadowing observed at
low $x$.  In order to evaluate these diagrams,
we will adopt a procedure of postulating photon-meson interactions, consistent 
with gauge invariance, and then verifying that there is no conflict with 
available information.

\vspace{0.5cm}
\begin{figure}[htb]
\begin{center}
\leavevmode
{\epsfbox{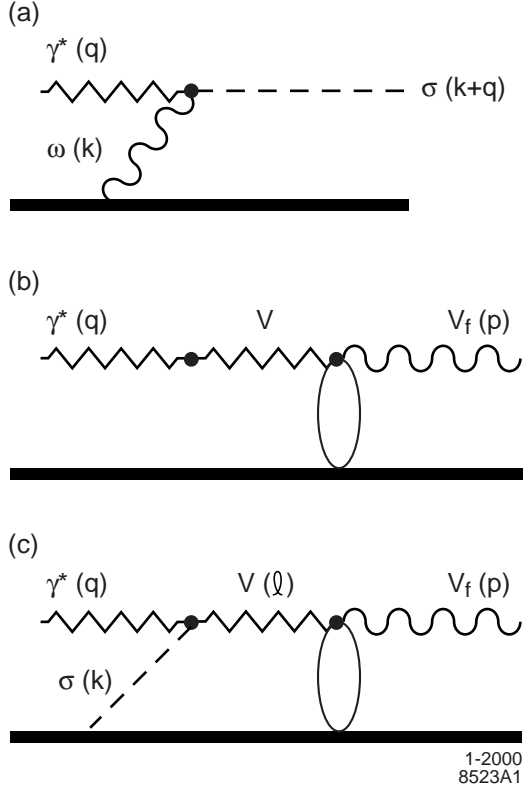}}
\end{center}
\caption[*]{\baselineskip 13pt
Diagrams for the high energy, low $Q^2\;\;\gamma^*$ nuclear scattering.  For
each mechanism, the
nuclear target is represented by a heavy line.  (a) Dominant contribution to
$\sigma_L(A)$; (b) Dominant contribution to the transverse cross section;
(c) Mesonic term that interferes with (b).  }
\label{hyperf1}
\end{figure}

Consider the longitudinal cross section.
We denote the contribution (per nucleon) to the nuclear hadronic tensor
caused by excess nuclear mesons as
$\delta W^{\mu\nu}$.  This is in addition to the gluonic effects which give a
non-zero value for $R$ for a free nucleon.
Then, using standard kinematic relations \cite{roberts},
we take the ratio of nuclear to nucleon values of the
longitudinal cross section to find
\be
{\sigma_L(A)\over \sigma_L(D)}=1+{Q^4\over\nu( \nu^2+Q^2)}
{\delta W^{00}\over F_2^D R_D}
(1+R_D).\label{kinrel}\ee
The notation $D$ represents the nucleonic value as
represented by the deuteron cross section, with
$ F_2^D$ and $ R_D$ taken from standard parameterizations of the
data\cite{f2ref,Abe}. The implication of Eq.~(\ref{kinrel}) is
that $\delta W^{00}$ must vary as $\nu^3$
to obtain a significant effect at small $x$.

In order to evaluate the effects of Fig.~1a,
we need to determine the $\gamma\omega\sigma$ interaction.
We postulate a gauge-invariant form
\bea {\cal L}_I=
{g\;e\over 2 m_\omega}\;F^{\mu\nu}(\omega_\nu\partial_\mu \sigma
-\omega_\mu\partial_\nu \sigma) \label{lios} \eea
where $F^{\mu\nu}$ is the photon field strength tensor.
In momentum space one can use
$ {\cal M}=
\ds {g\over m_\omega}
\left(p\cdot q\; \epsilon^\gamma\cdot\epsilon^\omega-p\cdot
 \epsilon^\gamma \;q\cdot\epsilon^\omega\right)
F_V(Q^2)$
in which $p_\mu$ is the momentum of the $\sigma$, and
we include a form factor $F_V$.  We seek
a constraint on the value of $g$ from the decay:
$\;\omega\to \sigma\gamma$.
The branching ratio for $\omega \to \pi^+ \pi^- \gamma <3.6\times 10^{-3}$
\cite{pdg},  which we assume to come from the process $\omega\to\gamma\sigma$
followed by the two pion decay of the $\sigma$ meson.
This process supplies a contribution to the $\omega$ width,
 $\delta\Gamma =(e^2/12\pi)(q_0^3g^2/ m_\omega^2),$
where
$q_0$ is the photon energy in the cm frame,
$q_0=(m_\omega^2-m_\sigma^2)/2m_\omega $
and $F_V(Q^2=0)=1$.
The predicted value of $\delta\Gamma$ and the extracted value of $g$ depend
strongly on the mass of the $\sigma$ meson, which is given \cite{pdg}
as the $f(400-1200)$.  If we use the average value of 800 MeV, the decay would
not occur, and the width of the $\omega$ would provide no constraint on the
value of $g$.  In these first calculations we choose $m_\sigma=600 $
MeV, and determine an upper limit for g:
 $ {g_{UL}^2\alpha}=.013 \approx 2\alpha.\label{nom} $
We shall take $g^2_{UL}=2$ as a nominal value.

The standard formula for a contribution to the hadronic tensor per nucleon is
\be
\delta W^{\mu\nu}={1\over 4\pi M_A}{1\over A}
\int d^4\xi\; e^{iq\cdot\xi}\;\langle
P|J^\mu(\xi)J^\nu(0)|P\rangle.
\label{start}\ee
The current $J^\mu$ is obtained from
 our interaction (\ref{lios}) using
 $J^\mu={\delta {\cal L}_I/ \delta A_\mu}$.
The state $|P\rangle$ is
the nuclear ground state, normalized as
$\hbox{$\langle P'| P \rangle$}
{=}\hbox{$2E(P)(2\pi)^3 \delta^{(3)}(P{-}P')$}$.
The only terms of $J_\nu$ we
need to keep are those which are
proportional to the large momentum of the outgoing $\sigma$ meson.
Keeping these
and evaluating Eq.~(\ref{start}),  gives the result:
\be
\delta W^{\mu\nu}={g^2F_V^2(Q^2)} 
\;{p\;\nu ^2\over 2 A}\; 
 {p^{\mu}p^{\nu}\over m_\omega^2}\int d\Omega_{\bf p} \;
 \omega^0(\bf{p-q})\omega^0(\bf{p-q}),
 \label{good}
\ee
in which negligible retardation effects in the $\omega$ propagator are ignored,
 $p^\nu$ is the momentum of the outgoing $\sigma $ meson, and $p$ is the
magnitude of its three-momentum, $\sqrt{\nu^2-m_\sigma^2}$. 
Note that only the time ($\mu=0$)
component of the field $\omega^\mu$ has a significant value.  The term
$\omega^0(\bf{p-q})$ is the Fourier transform of the nuclear vector
potential,
\be \omega^0({\bf{p-q}})=\int{d^3r\over (2\pi)^{3/2}}
e^{i(\bf{p-q})\cdot \bf{r}} \omega^0(\bf{r}).
\label{ftdef}\ee
which contains
the nuclear form factor.  The momentum transfer
{\bf{p-q}} will be huge on a scale of nuclear momenta $\sim 1/R_A$, and
the nuclear form factor nearly vanishes unless
${\bf p}$ is parallel to the direction of the
virtual
photon.  As a result, the angular integration gives
a factor $\sim 1 /(R_A^2\nu^2)$, which, when multiplied by
a large factor $\sim R_A^6$ arising from the product of two volume integrals
over the entire size of the nucleus,
leads to $\delta W^{00}\sim R_A\;\nu^3$.  This term
increases roughly as A$^{1/3}$, representing the net coherent effect of the
nuclear
$\omega$ field,  and it has the necessary dependence on $\nu$.

The essential input is $\omega^0(\bf{p-q})$.
For the $^{14}$N target, we use a Fermi form:
$ \omega^0_F(r)=V^0 /[ 1 + e^{(r-R)/a}].$ 
The resulting $\delta W^{00}_F$ is computed numerically using 
Eqs.~({\ref{good}},\ref{ftdef}).  The term $V^0$ is the central value 
of the field.
In mean field theory, $V^0$  is the vector potential for  the nucleon
divided by the $\omega$-nucleon coupling constant.  The value is
$V^0= 28$ MeV in nuclear matter (along with
the nucleon effective mass, $M^*=0.56M$) using
QHD1 \cite {bsjdw}.  However, the very small nucleon effective mass causes
the value of light cone plus momentum carried by nucleons to be far too small
for consistency with lepton-nuclear deep inelastic scattering data.
However, there are many versions of the theory.
One could include meson-meson interactions in the Lagrangian, as well as
RPA and Brueckner
correlations.  For example, in a mean field theory which includes $\sigma^3$
and $\sigma^4$ terms,
the  value  $V^0= 13$ MeV ($M^*=.84 M$)
gives a reasonable description of many nuclear properties
 \cite{fpw87}, and is also consistent
with nuclear deep inelastic scattering \cite{miller00}.  The
quark-meson coupling model \cite{qmc} also gives $M^*\simeq 0.8 M$.
Thus the expected  range of $V^0$ is \,$13\, <\,V^0<\, 28 $\, MeV.
In our present calculations we adopt the average value of $V^0=20$ MeV
as a nominal value.

In order to complete the description of our calculation we need to specify the
form factor.  The expression:
$ F_V(Q^2)=(1+Q^2/m_\rho^2)^{-1.5}$ is an analytic  representation
of the quark-loop diagram calculation of Ito and Gross \cite{ig}.
The results obtained with this form factor (labeled IG)
are shown in Fig.~2, using $R=2.89$ fm,
\hbox{$(g^2/ g^2_{UL})(V^0/20 \rm{MeV})^2=1$,} a value which
is  consistent with available information.
  To display the sensitivity to parameters, we also use a dipole form
factor (results labeled dipole in Fig.~2),
along with the values $(g^2/ g^2_{UL})(V^0/20 \rm{MeV})^2=1.2$.
In either case, the values of
$\sigma_L(A)$ are large, and a qualitative reproduction
of the HERMES data  is obtained. Clearly, the values of the couplings could be
much smaller; for example if  $(g^2/ g^2_{UL})=1/4$, then the computed ratio of
${\sigma_L(A)/\sigma_L(D)}$ would be $\sim 1.2$ for $x\sim 0.01$.

\begin{figure}
\unitlength1.cm
\begin{picture}(15,9)(-3,-10)
\includegraphics{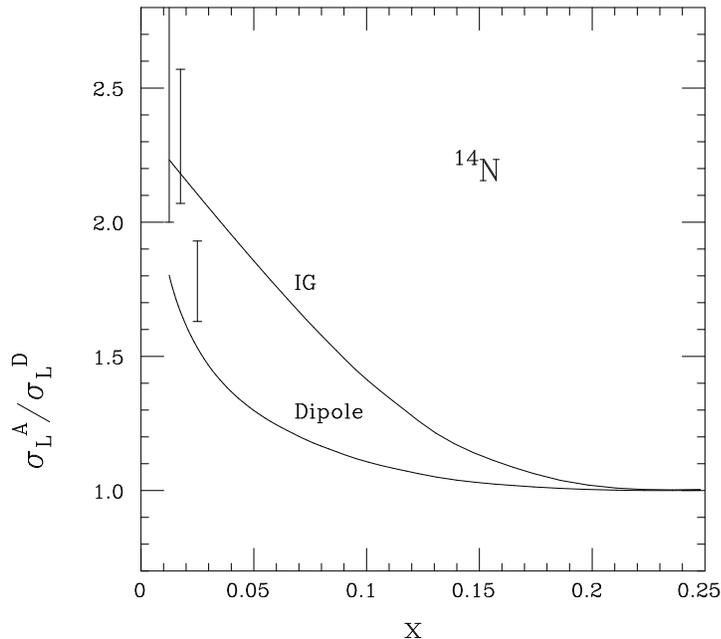}
\end{picture}
\caption{\baselineskip13pt ${\sigma_L(A)\over \sigma_L(D)}$, A=14, data from
 Ref.~\cite{Ackerstaff:1999ac}.  The labels IG,  dipole refer to
         form factors, see text.}
\label{fig:sla}
\end{figure}

We now return to the effects of the pion.  The $\gamma^*-\pi$ interaction
is represented by the current:
$(\partial^\mu\phi^*)\phi + \phi^*\partial^\mu\phi.$ 
Its use in Eq.~(\ref{kinrel}) gives the pionic contribution as:
\bea
{\sigma_L^\pi(A)\over \sigma_L(D)}&=&
{Q^4\over 2\nu^3}{1+R_D\over F_2^D R_D} \left[{\nu^2\over
 Q^2M_N}\right] f_{\pi/A}(x)F_\pi^2(Q^2) =
x f_{\pi/A}(x){1+R_D\over F_2^D R_D}F_\pi^2(Q^2).\label{second}\eea
The effects of the pion form factor are included via the term
$F^2_\pi(Q^2)$.  The pion distribution function $f_{\pi/A}(x)$ gives the 
probability that a nuclear pion has a plus momentum of $xM_N$.
As noted above,  $f_{\pi/A}(x)$ vanishes as $x$ approaches zero.
The value of $\sigma_L^\pi(A)/ \sigma_L(D)$ for the data point at $x=0.0125$ is
estimated using \cite{f2ref,Abe}
$F_2^D(x)=0.22$,  $R_D=0.36$,  and $f_{\pi/A}(x)\approx 0.01$ from the
 calculation of Ref.~\cite{jm} for infinite nuclear matter.
This calculation uses a version of the Ericson-Thomas \cite{et}
model in which the parameters are chosen  to be consistent with nuclear
deep inelastic and Drell-Yan data.  The result is
$\sigma_L^\pi(A)/ \sigma_L(D) \sim 0.01$ which is negligible.  The
function $f_{\pi/A}(x)$ peaks at $x=0.25$ with a value (for the charged pions)
of 0.24.  For this $x$, $F_2^D\approx 0.23,\;R_D=0.3$.  These values give
$\sigma_L^\pi(A)/ \sigma_L(D)\approx  F_\pi^2(Q^2).$
For the kinematics in the HERMES experiment,
$Q^2\approx 3.1$ GeV$^2$ and the square of the form factor is $\approx 0.03$.
Thus one obtains
a contribution of about 0.03.  This is small compared to the vector meson
current contributions, but it is not entirely negligible.  However,
in an experiment at $x=0.25$ with $Q^2\sim 1\;{\rm GeV}^2$,
nuclear-coherent pions would contribute significantly to $\sigma_L$.

The presence of nuclear pions opens the door to a
variety of new experiments. 
One could study the exclusive final state of
pionic reactions such as
$\gamma^*\,\,{}^3\hbox{He} \to \pi^+ ~{}^3\hbox{H}$,
 $\gamma^* ~^{14}\hbox{N}\to \pi^- ~
^{14}\hbox{O}$, and $\gamma^* ~^{14}\hbox{N} \to \pi^+ ~^{14}\hbox{C}$,
where the final state nuclides can
be formed in their ground or excited states. These
nuclear-diffractive electroproduction processes
can be used to test predictions from QCD for the $Q^2 $
and $t$ dependence of the off-shell mesonic form factors $F_M(Q^2,t)$.
Conversely, by tagging the nuclear final state in an inelastic reaction,
one could identify the structure functions of off-shell mesons, as in the
Sullivan process.  Such processes are analogous to the measurements of
the pomeron structure function in diffractive deep inelastic scattering.

Next we examine the nuclear value of the transverse cross section $\sigma_T$.
Our explanation of the data requires
a significant destructive interference effect at low $Q^2\approx 0.5-2\;
{\rm GeV}^2$, which decreases rapidly as $Q^2$ increases.  Furthermore,
the shadowing of the real photon ($Q^2=0$) is not very strong, and it is well
explained by conventional vector meson dominance models \cite{bauer}.
Thus consistency with all available data demands an amplitude for
$\gamma^*\sigma\to V$ which  vanishes, or is small,
as the $Q^2$ of the virtual photon $\gamma^*$ approaches 0.  This means that
 measuring the real photon decays of the vector mesons
provides no constraints on the coupling constant.
We postulate a sum of two gauge-invariant forms:
 \bea
 {\cal L}_{\rho\gamma\sigma} &=& {g_{\rho\gamma\sigma}\over 2m_\sigma}
 \left[F_{\mu\nu}\rho^{\mu\nu}\sigma
 + \lambda
 F_{\mu\nu}(\rho^\mu\partial^\nu\sigma-\rho^\nu\partial^\mu\sigma)\right]
 \label{vo1} \\
&  \to& {-g_{\rho\gamma\sigma}\over
    m_\sigma}\left[(q\cdot(k+q)\epsilon^\gamma\cdot\epsilon^\rho-
    q\cdot\epsilon^\rho k\cdot\epsilon^\gamma)
+\lambda( q\cdot\epsilon^\rho k\cdot\epsilon^\gamma
-q\cdot k\epsilon^\gamma\cdot\epsilon^\rho) \right].
    \eea
 The choice $\lambda=1$ leads to an effective Lagrangian
$ {g_{\rho\gamma\sigma }\over m_\sigma}
 \left( Q^2\epsilon^\gamma\cdot\epsilon^\rho+q\cdot
 \epsilon^\gamma\;q\cdot \epsilon^\rho\right)F_V(Q^2),$
 which is the form that we adopt.
 We choose a gauge such that $ q\cdot \epsilon^\gamma=0.$
 Note also that the difference in the two terms of Eq.~(\ref{vo1}) is due to
 the source of the electromagnetic field, so that $ {\cal L}_{\rho\gamma\sigma}$
 depends on the virtuality  of the photon.  A form factor $F_V(Q^2)$ is 
introduced.
 We simplify the calculation by considering only one intermediate
 vector meson state,
 which we label $\rho$.  We shall treat its mass denoted $m_\rho$, which can
 range from the mass of the physical $\rho$ meson up to values comparable
 with $\nu\sim$ 20 GeV here, as a free
 parameter.
 Then the usually  dominant
 term ${\cal M}_{\rm dom}$ of Fig.~1b takes the form:
 \bea
 {\cal M}_{\rm dom}=e{m_\rho^2\over f_\rho}{-1\over Q^2+m_\rho^2}\,
 \epsilon^\gamma \cdot\epsilon^\rho\int d^4x'\;T_A(x')e^{i(p-q)\cdot x'},
 \label{dom}
 \eea
 in which $p$ is the momentum of the final vector meson,  and $T_A(x)$ 
represents
 the purely imaginary final-state interaction with the target nucleus which
 converts the intermediate vector meson to the final  vector meson.

 The nuclear-coherent term of Fig.~1c  takes the form
\bea
 {\cal M}_{\sigma}= Q^2 F_V(Q^2)\epsilon^\gamma \cdot\epsilon^\rho
 {g_{\rho\gamma\sigma }\over m_\sigma}
 \int d^4x\int\;d^4x'\int{d^4 l\over (2\pi)^4}e^{ip\cdot x'}T_A(x')
 {e^{-il\cdot(x'-x)}\over l^2-m_\rho^2+i\epsilon}
e^{-iq\cdot x}\sigma(x).\eea
The evaluation is straightforward.
The integration over $d^4x$ gives a delta function setting $l$ equal to $k+q$.
The propagator contains a factor $(k+q)^2-m_\rho^2+i\epsilon\approx
-2\nu k^3-Q^2-m_\rho^2+i\epsilon$
in which the direction of the photon momentum
is taken as the positive $z(3)$ axis.
The approximation of neglecting $k^2(\ll m_\rho^2)$
yields a propagator of the eikonal form, so that after integration
\bea
 {\cal M}_{\sigma}
= {-iQ^2F_V(Q^2)\over 2\nu} \epsilon^\gamma \cdot\epsilon^\rho
 {g_{\rho\gamma\sigma }\over m_\sigma}\int d^4x'e^{i(p-q)\cdot x'}T_A(x')
 \int_{-\infty}^{z'} dz
 e^{-i{Q^2+m_\rho^2\over 2\nu}({z}'-z)}\sigma({\bf x_\perp'},z)
\label{stf}\eea
The overall factor of $i$ arises from the eikonal propagator, and another
phase appears in the exponential.  This phase factor vanishes for extremely
large values of $\nu$, but it is important here because the
large nuclear
radius enters ($z'-z\sim R_A$).  Thus
$ {\cal M}_{\sigma}$ can interfere with $ {\cal M}_{\rm dom}$.
Note that for large enough values of $\nu$,
$ {\cal M}_{\sigma}\sim  ({Q^2+m_\rho^2)/ 2\nu}R_A.$ This gives a
rough guide to the dependence on $Q^2 $ and $A$.
Thus there is a very strong dependence on the specific  value of $m_\rho^2$
used. We choose this parameter to be 5 times the square of the
mass of the physical $\rho$ meson.
The amplitude $ {\cal M}_{\sigma}$ is proportional to
the central value of the nuclear $\sigma$ field
$\sigma(0)$ which takes on a value of ${-}43$
MeV in QHD1.  If the non-linear model \cite{fpw87} is used, $\sigma(0)={-}29$
MeV.  In our present calculations we choose the average value $\sigma(0)={-}36$
MeV.
The results shown in Fig.~3 are obtained by numerical integration and assuming
that the time dependence of $T_A(x')$ is the same for all positions within the
nucleus.  The  coupling constants and form factors are taken as
$g_{\rho\gamma\sigma}=$2.4, 2.9, and
$F_V(Q^2)=\exp\left[-(Q^2-({\tilde m}_\rho)^2)R_V^2/6\right]$,
with $R_V=0.99$ fm.  We use choose ${\tilde m}_\rho$ to be
the physical mass of the $\rho $ meson. This  defines the value of
$F_V(0)$, and is  simply a convention for defining the value of
$g_{\rho\gamma\sigma}$. 
An exponential, rather than power-law fall-off, is
essential in reproducing the data.  This form factor implies a strong $Q^2$
dependence at fixed values of $x$. The parameter $R_V$ takes on the value of
a typical hadronic size.  

How large should $g_{\rho\gamma\sigma}$ be?  We
can compare   the strength of the $\gamma\to\rho\sigma $
transition with that of $\gamma\to\rho$ at $Q^2=m_\rho^2$:
$g_{\rho\gamma\sigma} s{\sigma/ m_\sigma}\ vs. \ {e / f_\rho}$.
The factor  $\sigma$ is  expected from the $\sigma$ model
to be of the order of $M/g_{\pi NN}\approx 70$ MeV. Thus the couplings are
comparable if
$g_{\rho\gamma\sigma}\approx 0.5$. This is about ten times smaller than the
coupling needed to describe the strong  shadowing
observed by HERMES. However, $g_{\rho\gamma\sigma}=$ 2.9
is not ruled out by any existing data. On the other hand,  if
$g_{\rho\gamma\sigma}=0.5$,
${\sigma_T(A)/ \sigma_T(D)}\approx 0.95$ for $x\sim 0.01$.
\begin{figure}
\unitlength1.cm
\begin{picture}(15,9)(-3,-10)
 \includegraphics{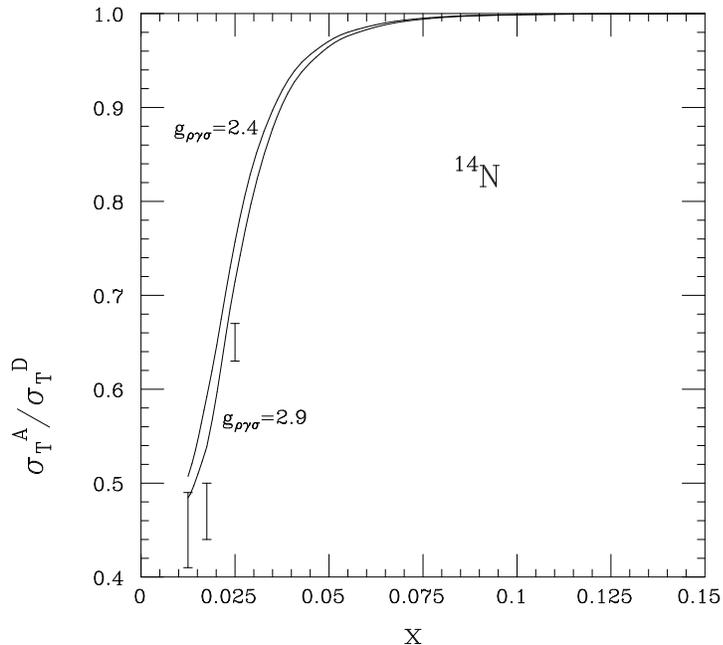}\hspace{1in}
\end{picture}
\caption{${\sigma_T(A)/ \sigma_T(D)}$, A=14, data of
 Ref.~\cite{Ackerstaff:1999ac} }
\label{fig:sta}
\end{figure}
 The nuclear enhancement of $R$ is obtained from
computing the
ratio of the results of Eqs.~(\ref{good},\ref{stf}).  This is shown in Fig.~4,
where it is seen one has a reasonably good description of the data.
\begin{figure}
\unitlength1.cm
\begin{picture}(15,9)(-3,-10)
 \includegraphics{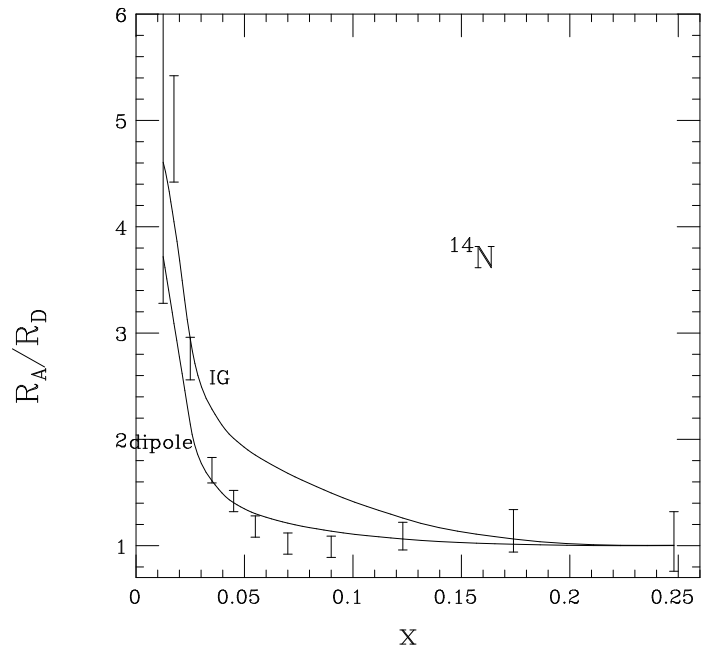}\hspace{1in}
\end{picture}
\caption{${R(A)/ R(D)}$, A=14 }
\label{fig:ra}
\end{figure}

Data also exist for the $^3$He target.  To address these data properly, one
should perform a
three-body calculation.  We do not attempt this here.  To understand if our
theory has a reasonable dependence on $A$, we simply rescale the radius
parameter
$R$ by a factor $(3/14)^{1/3}$ and take the number of nucleons to be 3.
The result, as shown in Fig.~5, is in qualitative agreement with the HERMES
$^3$He data.
\begin{figure}
\unitlength1.cm
\begin{picture}(15,9)(-3,-10)
 \includegraphics{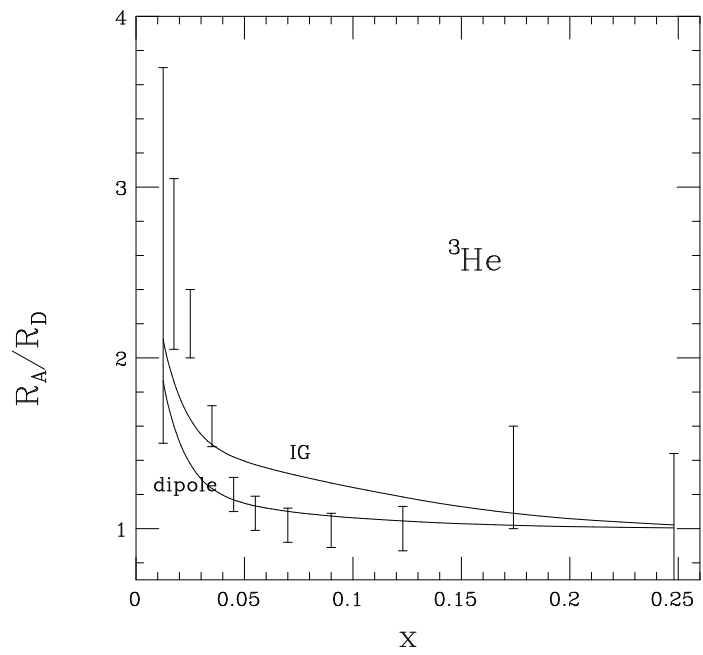}\hspace{1in}
\end{picture}
\caption{${R(A)/R(D)}$, A=3,
 data from
 Ref.~\cite{Ackerstaff:1999ac}}
\label{fig:ra3}
\end{figure}

Our analysis shows that it is possible, with reasonable coupling strengths, to
reproduce the salient features of the HERMES data. The  calculations employ a
particular choice of couplings, optimized to reproduce the HERMES, without
contradicting other experimental constraints.   For example, the strengths 
of the meson
fields in nuclear medium which we use are at least roughly consistent with
measured nuclear binding energies, nuclear densities, and nuclear deep
inelastic and Drell-Yan data.  However, the constants $g,g_{\rho\gamma\sigma}$
have never been measured, and their magnitudes could turn out to be
small.  The increase in the longitudinal cross section is readily explained by
the exclusive
process $eA\to e'\sigma A$ via $\omega$ exchange. The nuclear enhancement
follows from the coherence of the $\omega$ field. The strong
shadowing of $\sigma_T$
at small $Q^2$ requires a special choice of the effective Lagrangian
which is theoretically and empirically allowed, but does not seem to follow from
any general principle.  Thus, more conservatively, we cannot
rule out the HERMES data using our theory.

Further tests of our model are possible.  An immediate consequence
would be the observation of  exclusive mesonic states
in the current fragmentation  region.  In particular, our description of
$\sigma_L(A)$ implies significant nuclear-coherent production of
$\sigma$ mesons along the virtual photon direction.  Our model for
the strong shadowing of coherent meson effects in $\sigma_T(A)$ can be tested
by measurements performed at the same value of $x$ but different values of
$Q^2$ than HERMES used.

The prospect that the mesonic fields which are responsible for nuclear binding
can be directly confirmed as effective fundamental constituents of nuclei at
small $x$ and  $Q^2\sim 1\;{\rm GeV}^2$
is an exciting development at the interface of traditional nuclear physics
and QCD.  The empirical confirmation of
nuclear-coherent meson contributions in the final state would
allow the identification of a specific dynamical mechanism for higher-twist
processes in electroproduction.
Clearly, these concepts should be explored further, both
experimentally and theoretically.

\section*{Acknowledgments}

This work was supported in part by the United States
Department of Energy under contract numbers DE--AC03--76SF00515,
and DE-FG03-97ER4104 and by a
grant from the U.S.-Israel Binational Science Foundation.
We appreciate the hospitality of the CSSM in Adelaide where some of this
work was performed.  We thank  W.~Melnitchouk, S. Rock,
M. Strikman,  and especially  G. van der Steenhoven, for useful discussions.

\end{document}